\RequirePackage{fix-cm}
\documentclass[smallextended]{svjour3}       
\usepackage{makeidx}         
\usepackage{graphicx}        

\graphicspath{{images/eps/}{images/pdf/}} 
\smartqed  
\usepackage{graphicx}
\usepackage{multirow}
\usepackage{lscape}
\usepackage{colortbl}
\usepackage{eqlist}
\usepackage{cellspace}
\usepackage{longtable}

\usepackage{amssymb}
\newcommand{\rouble}{{\rm{P}\kern-.635em\rule[.5ex]{.52em}{.04em}\kern.11em}}

\begin{document}

\title{Nonparametric modeling cash flows of insurance company 
}

\titlerunning{Nonparametric modeling cash flows of insurance company }        

\author{ Valery Baskakov, Nikolay Sheparnev \& Evgeny Yanenko
}


\institute{International Actuarial Advisory Company (IAAC), Llc Member of European Actuarial \&
Consulting Services (EURACS), Malaya Filevskaya st. 50, ap. 80, Moscow 121433, Russia
           \at  \at
           V. Baskakov \at
              Tel.: +7-903-100-2660\\
              \email{chief@actuaries.ru}           
           \and
           N. Sheparnev \at
              Tel.: +7-915-468-9315\\
              \email{sheparnevnv@iaac.ru}
           \and
           E. Yanenko \at
              Tel.: +7-916-407-9423\\
              \email{yanenko@iaac.ru}
}

\date{Received: 12-04-2019 / Accepted: date}

\maketitle

\begin{abstract}
The paper proposes an original methodology for constructing quantitative statistical models based on multidimensional distribution functions constructed on the basis of the insurance companies' data on inshurance policies (including policies with deductible) and claims incurred. Real data of some Russian insurance companies on non-life insurance contracts illustrate some opportunities of the proposed approach. The point and interval estimates of net premium, claims frequency, claims reserves including IBNR and OCR, are thus obtained. The resulting estimate of claims reserves falls in the range of reasonable estimates calculated on the basis of traditional reserving methods (the chain-ladder method, the frequency-severity method and the Bornhuetter-Ferguson method).

The proposed methodology is based on additive estimates of a company's financial indicators, in the sense that they are calculated as a sum of estimates built separately for each element of the sample (claim). This allows using the proposed methodology to model insurance companies' financial flows and, in particular, to solve the problems of reserve redistribution between particular segments of insurance portfolio and/or time intervals; to adjust risk as part of financial reporting under IAS 17 Insurance Contracts; and to deal with many other tasks.

The accuracy of insurance companies' financial parameters estimate based on the proposed methods was tested by statistical modeling. IBNR was used as the test parameter. The modeling results showed a satisfactory accuracy of the proposed reserve estimates.

\end{abstract}

\section{Introduction}
\label{intro}


The development of quantitative statistical models used in modeling insurance companies' cash flows usually implies the knowledge of the distribution functions of random variables involved in this process. For example, if the distribution function $\textbf{F} (s)$ of claims s which includes zero claims is known, then the net premium $T_n$ is equal to the expected claim value, i.e.
$$
    T_n=\textbf{M} (s)=\int\limits_0^\infty s \cdot d \textbf{F} (s).
$$
If we additionally know the expected number of claims incurred but not reported $m$ and their average value $\textbf{M}^* (s)$, then the IBNR reserve is
\begin{equation}
\label{equ:1}
    R=m \cdot\textbf{M}^* (s).
\end{equation}
Note that generally $\textbf{M}(s) \ne \textbf{M}^* (s)$.

It is evident that knowing the corresponding probability distributions allows to develop similar formulas for calculating incurred but not paid claims; to build distributions of the corresponding reserves over time, i.e. to estimate the cash flows necessary to settle the claims timely; to adjust risk as part of financial reporting under IAS 17 Insurance Contracts; and to deal with many other tasks. However, in practice the structure of available data is often complex and does not allow to use traditional statistical methods for estimating the necessary probability distributions.

In fact, the current insurance statistics cannot present a complete-data system even in theory. Indeed, insurance is a process evolving over time. Therefore, at any moment of actuarial calculations a company's portfolio always contains both completed and incomplete policies with insured claims partly reported and settled, and partly not yet reported and/or reported but not settled. This is the main reason (among many others) to censor observations both in time and in claim value. For example, both property and casualty insurance policies contain deductibles and/or limits, which also leads to the collection of incomplete data. The deductible prevents from capturing data on claims below a certain fixed amount of money (left censoring), and the limit prevents from capturing the exact amount of claim (right censoring). The further reading provides additional factors of the truncating and/or censoring of insurance statistics data  (for example see~\cite{Baskakov2019}, ~\cite{Klein2003}).

This paper applies the ideas presented in one of the authors' works ~\cite{Baskakova2014}, ~\cite{Baskakov2010}   and generalizes them for the case of insurance policies with deductibles, applying qED estimate of the multivariate distribution function for censored data ~\cite{Baskakov1996}. The paper is arranged as follows. In Section~\ref{sec:1} we discuss the procedure for collecting insurance statistics and provides a formal description of a censored sample structure. In Section~\ref{sec:2} we outline the procedure for building a qED estimate of the joint distribution function of both the claim and time interval between the dates of an insurance claim and its reporting. In Section~\ref{sec:3} we discuss methods for solving a number of applied problems based on historical data, including an estimate of insurance rates and reserves. The estimations are compared with similar estimates obtained by traditional reserving methods. In Section~\ref{sec:4} we present the results of the accuracy simulation study of the two-dimensional censored data estimation.

\section{Statistical data structure and censored data sampling}
\label{sec:1}

The main source of insurance statistics in the Russian Federation is the Register of Insurance (Co-Insurance) Policies and the Register of Claims and Early Termination Insurance (Co-Insurance) Policies. The Registers' formats and entry procedures are set in the insurer's internal documents, however it is the regulator (the Bank of Russia)\footnote{Regulations on the insurance reserves buildup for insurance other than life insurance Authorized by decision of the Bank of Russia No. 558-P of 16.11.2016 Registered in the Ministry of Justice of the Russian Federation under No.45054 on 29.12.2016}  that specifies the compulsory data to be entered and kept by an insurer for at least 5 years from the date of the full performance of obligations under the policy.

The Registers shall contain the following information:
\begin{enumerate}
\item[--] Policy No.

\item[--]  Effective date of the policy

\item[--]  Date of liability commencement

\item[--]  Policy period

\item[--]  Liability period(s), if other than the policy period

\item[--]  Sum(s) insured

\item[--]  Policy early termination (cancellation) date

\item[--] Policy amendment date(s)

\item[--] Insurance claim(s) report date(s)

\item[--] Insurance claim(s) occurrence date(s)

\item[--] Insurance claim ID

\item[--] Reported claim amount(s) as well as information on change in the date(s) and amount(s) of the reported claim(s) during its settlement, specifying the insurance claim ID

\item[--] Insurance payment date(s) specifying the insurance claim ID

\item[--] Insurance payment amount(s) specifying the insurance claim ID, etc.
\end{enumerate}

To proceed further, we should clearly see the structure of available insurance statistics. Therefore we provide its formal description as the next step. We are considering the cases when an insurer keeps statistical records in accordance with the Russian law, i.e. the records are kept separately for each policy, and the following indicators (among others) are recorded:

\begin{description}
\item[$t_i^1$] ~is the start date of the insurance policy No. $i$, $i=1, \dots ,n$ (where $n$ is the number of policies concluded, and $i$ hereinafter refers to the policy number)

\item[$t_i^2$]  ~is the policy early termination (cancellation) date

\item[$\tau_{ik}^1$]  is the occurrence date of the insurance claim No. $k$

\item[$\tau_{ik}^2$] is the report date of the insurance claim No. $k$

\item[$\tau_{ik}^3$] is the settlement date of the insurance claim No. $k$

\item[$t~~$] is the date of insurance statistics collection (reporting date)

\item[$s_{ik} (t)$] is the total amount of payments relating to the insurance claim No. $k$ on date $t$.
\end{description}

A formal criterion is assumed to be in place to consider the claim paid at the date $t$ as settled, i.e. to ensure that no additional payments are due in the future or otherwise $s_{ik} (t)=s_{ik} (\infty)=S_{ik}$. Further on, we are using the indicator $\delta_{ik}$ which takes on values as follows: 0 - claim settled; 1 - claim reported but not settled; 2 - claim incurred but not reported.

In some cases, additional options are included as clauses in an insurance policy, for example deductibles $d$, that determine how much of an insurance-covered expense is borne by the policyholder. Or the policy determines the limitation period, i.e. the time $L$ for the policyholder to report the claim to the insurer (under the Insurance Rules); or sets liability limit for each individual insurance claim. We will take the abovementioned possibilities into consideration. Further on, if the policyholder did not report insurance claim No. $k$ within the limitation period, i.e. if $t-\tau_{ik}^1>L$, then we assume there were no insurance claims during this period or, equivalently, an insurance claim $S_{ik}=0$ occurred and was settled $(\delta_{ik}=0)$, and the period between claim reporting and occurrence date is $\tau_{ik}=\tau_{ik}^2-\tau_{ik}^1=\infty$. Obviously, these provisions do not limit the reasoning in general.

Let us set $[t_1, t_2 )$ - the time interval for which the financial flows (claims) are calculated. Let the indicated dates be interrelated as follows:
\begin{equation}
\label{equ:2}
 \begin{array}{cl}
     & t_i^1<t_i^2;  \\
     & t>t_i^1; 		 \\				 			
     & \tau_{ik}^1 \le \min(t,t_i^2 ); \\
     & t \le t_1<t_2. \\
 \end{array}
\end{equation}

\begin{figure}
\center
\begin{picture}(340,140)(0,-10)

\put(0,15){\line(1,0){335}}
\put(20,11){$\shortmid$}
\put(160,11){$\shortmid$}
\put(300,11){$\shortmid$}
\put(0,0){01.01.2017}
\put(140,0){01.01.2018}
\put(280,0){01.01.2019}
\put(313,125){Policy}
\qbezier[59](310,15)(310,65)(310,130)

\put(210,40){\line(1,0){105}} \put(210,40){\circle*{4}}
\put(210,30){$t_4^1$}
\put(115,65){\line(1,0){140}} \put(115,65){\circle*{4}} \put(255,65){\circle*{4}} \qbezier[59](255,65)(300,65)(315,65)
\put(115,55){$t_3^1$} \put(255,55){$t_3^2$}
\put(155,55){$\tau_{3,1}^1$} \put(155,63){$\shortmid$}
\put(190,55){$\tau_{3,1}^2$} \put(190,63){$\shortmid$}
\put(215,55){$\tau_{3,2}^1$} \put(215,63){$\shortmid$}
\put(285,55){$\tau_{3,2}^2$} \put(285,63){$\shortmid$}

\put(320,113){1}\put(320,88){2}\put(320,63){3}\put(320,37){4}

\put(50,90){\line(1,0){80}} \put(50,90){\circle*{4}} \put(130,90){\circle*{4}} \qbezier[140](130,90)(300,90)(315,90)
\put(50,80){$t_2^1$} \put(130,80){$t_2^2$}
\put(75,80){$\tau_{2,1}^1$} \put(75,88){$\shortmid$}
\put(105,80){$\tau_{2,1}^2$} \put(105,88){$\shortmid$}
\put(245,80){$\tau_{2,1}^3$} \put(245,88){$\shortmid$}

\put(0,115){\line(1,0){140}} \put(0,115){\circle*{4}} \put(140,115){\circle*{4}} \qbezier[134](140,115)(300,115)(315,115)
\put(0,105){$t_1^1$} \put(140,105){$t_1^2$}
\put(280,105){$t_1^2+L$} \put(280,113){$\shortmid$}

\end{picture}

\caption{Insurance statistics collection diagram}
\label{fig:1}       
\end{figure}
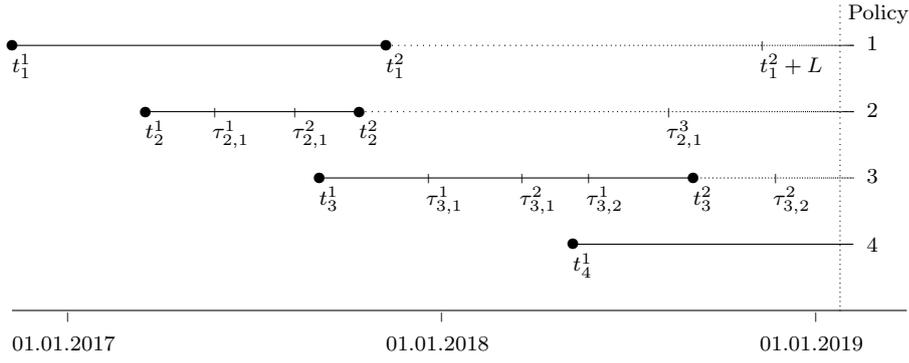

In general, insurance policies start at random points of time $t_i^1, i=1, \dots ,n$,
therefore the period $t-t_i^1$ from the start of policy No. $i$ to current date $t$ is also random. As a result, individual insurance policies cease to be under observation at different stages of their development, for example, within the insurance policy period (Figure~\ref{fig:1}, policy  4) or quite a long time after its termination (Figure~\ref{fig:1}, policy  1). Therefore, the actual claim amount $S_{ik}$ for the insurance claim\footnote{It is necessary to distinguish the actual claim $S_{ik}$ against insurance event No. $k$ from the claim  $s_{ik}$ against the same event paid on a specific date $t$. These values are obviously interrelated by the ratio $S_{ik} \ge s_{ik} (t)$, and the equality is achieved if the claim is considered settled based on the formal criterion}  is not always precisely known, and, in some cases, the very fact of insurance claim occurrence cannot be reliably established due to the time lag between the claim occurrence and report dates. If an insurance policy envisages a deductible $d_i$, then this is actually the same as that the claims $S_{ik} < d_i$ against the policy $i$ are not reported at all.

Let us consider the information about the actual claim $S$ obtainable from the available insurance statistics. Note that insurance policies may have different validity periods (for example, Figure~\ref{fig:1}, policy  2), and for some of them more than one insurance claim may occur (Figure~\ref{fig:1}, policy  3). In order not to exclude these possible cases, let us use the following method. Let us divide the effective period of insurance policy No. $i,~i=1, \dots ,n$ into equal time intervals $\Delta t_k, k=1,\dots ,n_i$ such that during $\Delta t_k$ one and only one insurance claim occurs, including zero claims $S=0$. For a number of insurance types an interval of 24 hours\footnote{By default, let us consider the time interval $\Delta t_k$ as 24 hours, if not defined otherwise}  may be used as $\Delta t_k$.

Let us assume that in case of zero claim, the time between the occurrence and report dates is infinity, i.e. $\tau=\infty$. We see that in practice this coincides with a situation where no claim occurred, but this convention is needed to simplify the subsequent reasoning. In this particular claim the expected claim is $\textbf{M}(s) \equiv 0$.

Let us move from considering individual insurance policies to the analysis of the aggregate of $n=\sum_i n_i$  days\footnote{Hereinafter $n$ means the sample scope equal to the number of policies sold or to the aggregate number of all policies' validity days, and this does not lead to ambiguity since it is always clear from the context which value is involved}. Given the assumption that every day one and only one insurance claim occurs per each insurance policy, then each day preceding the reporting date $t$ is also the date of the claim (including incidents with zero claims). Therefore, when using consecutive numbering (for policies) for $\tau_{ik}^1,k=1,\dots ,n$, index $i$ indicating the policy number may be omitted. The same provision applies to the values $t$, $t^1$ and $t^2$, represented in the coordinate system tied to the claim date:
\begin{equation}
\label{equ:3}
 \begin{array}{cl}
     & t_k=t-\tau_{ik}^1;  \\
     & t_{1k}=t^1-\tau_{ik}^1;		 \\				 			
     & t_{2k}=t^2-\tau_{ik}^1. \\
 \end{array}
\end{equation}

\begin{figure}
\center
\begin{picture}(300,100)(0,-5)
\put(150,50){a)}
\put(150,0){b)}
\put(80,15){\line(1,0){220}}
\put(80,13){$\shortmid$}
\put(180,13){$\shortmid$}
\put(240,13){$\shortmid$}
\put(280,13){$\shortmid$}
\put(80,20){0}
\put(180,20){$t_k$}
\put(240,20){$t_{1k}$}
\put(280,20){$t_{2k}$}

\put(0,65){\line(1,0){300}}
\put(0,65){\circle*{4}}
\put(0,70){$t_i^1$}
\put(80,70){$\tau_{ik}^1$}
\put(80,63){$\shortmid$}
\put(150,65){\circle*{4}}
\put(150,70){$t_i^2$}
\put(180,70){$t$}
\put(180,63){$\shortmid$}
\put(240,70){$t_1$}
\put(240,63){$\shortmid$}
\put(280,70){$t_2$}
\put(280,63){$\shortmid$}

\end{picture}

\caption{Statistical data structure}
\label{fig:2}       
\end{figure}
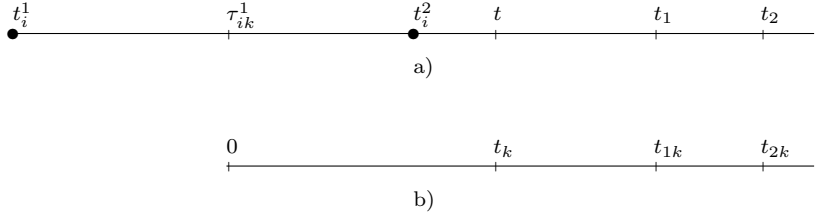

Thus, the boundaries of time interval $[t_1, t_2)$ fixed in the coordinate system of calendar time defined by formulas~(\ref{equ:3}), depend on the claim date and are equal to $[t_{1k}, t_{2k})$. The considered situation is visualized in Figure~\ref{fig:2}.

Under the considered data collection scheme, the claims (including zero claims) occurred in the interval $\Delta t_k$ are not represented by the actual values $S_k$, but by some value sets $C_k^s$ of possible claim values called censoring sets such that
\begin{equation}
\label{equ:4}
   S_k \in C_k^s, k=1,\dots ,n,	
\end{equation}
where $n=\sum_i n_i$.

It should also be noted that the structure and size of the censoring sets $C_k^s$ depend on the current date $t$. Indeed, the analysis of the insurance statistics collection scheme shows that for insurance claims declared at the time $\tau_k^2<t$ information is available regarding the occurrence time $\tau_k^1$ and the amount of the claim paid $s_k (t)$, as well as its status: settled $S_k=s_k (t)$ and $C_k^s=\{S_k \}$ or outstanding, $S_k \ge s_k (t)$ and $C_k^s=\{s:s \ge s_k (t)\}$. If, at the moment $\tau_k^1<t$, the insurance claim has not yet been reported, then only the following can be concluded based on the data of the scheme:
\begin{equation}
\label{equ:5}
   S_k \ge 0~~ or~~  C_k^s=\{s:s \ge 0\}.
\end{equation}

We see that formula~(\ref{equ:5}) contains no valuable information about the outstanding claims since exactly the same had been known before the policy was sold (the censoring set $C_k^s$ coincides with the claim definition domain $s$).

However, the situation in question can be resolved. The point here is that claims occur and develop in time, which is clearly seen in (Figure~\ref{fig:1}). Comparing policy  4 and policy  1, it is obvious that they provide different information about the claims that may have already occurred before the current date $t$. As for the first case, the policyholder can still report on claims that occurred at the moment $\tau_{4k}^1 \in \Delta t_k, k=1, \dots , n_4$ on any day $\tau_{4k}^2 \in (t, \tau_{4k}^1+L]$; however, in case of policy  1 this is not possible (because $\tau_{1k}^1+L<t$ is true for every $\tau_{1k}^1 \in t_k, k=1, \dots , n_1)$, and according to our provisions we can assume that the claim amount is $S_{1k}=0$, and the interval between the claim occurrence and report dates is $\tau_{1k}= \tau_{1k}^2-\tau_{1k}^1=\infty$.

Thus, some information about IBNR is still found in the statistics collected by insurers, however it is not available while analyzing the sampled random value $S_k,~ k=1, \dots ,n$. This information may be revealed, inter alia, through 3D random vector analysis.
\begin{equation}
\label{equ:6}
  (S_k, \tau_k, t_k), ~~ k=1,\dots , n,						
\end{equation}
where $\tau_k=\tau_k^2-\tau_k^1$ is the period between the occurrence and report dates, and $t_k=t-\tau_k^1$ is the period from the occurrence date until the present date. We remind here that according to our provisions, $\tau_k^1$ is the current 'policy' date, consistently taking on all values from the intervals $[t_i^1,min(t_i^2,t)], i=1, \dots , n$, including days with zero claims and claims not reported.

The vector sampling scheme~(\ref{equ:6}) has already been considered above. Obviously, this sample as well as sample~(\ref{equ:4}) is censored, with the first two coordinates $s$ and $\tau$ being censored, and the information about the coordinate $t$ is complete. The coordinates $s$ censoring the sets $C_k^s$ are defined earlier, and the sets $C_k^s$ may only be of two types: $C_k^{\tau}=\{\tau : \tau > t - \tau_k^1 \}$, if claim No. $k$ has not yet been reported and the insurance validity period has not yet expired or, otherwise, $C_k^{\tau}=\{\tau_k\}$ (including $C_k^{\tau}=\{\infty\}$ at $t-\tau_k^1 > L)$.

In the data collection scheme considered, the censoring value is represented by the time $t$ or the associated random variable\footnote{Hereinafter, a simple $t$ represents the value $t-\tau_k^1$. This does not lead to ambiguity since the intended value is always clear from the context}  $t_k=t-\tau_k^1$ which is precisely determined by the insurance company and is not dependent statistically on the values $s$ and $\tau$. This makes it possible to simplify the estimation of marginal claim distribution function $\textbf{F} (s)$ without corrupting the generality, by confining ourselves to the construction of the 2D distribution function $\textbf{F} (s, \tau)$  since in view of the abovementioned independence of the individual vector $(s, \tau, t)$ coordinates, the equality $\textbf{F} (s, \tau, t) =\textbf{F} (s, \tau) \cdot \textbf{F}(t)$ is valid.

Thus, according to the insurance company, the random vector $(s, \tau)$ is represented by the censored set
$$
 C_k \equiv  C_k^s \times C_k^{\tau} \ni (S_k,\tau_k), ~k=1, \dots , n	
$$
i.e. the set of possible values of the actualized vector $(s, \tau)$, which (the set) is actually seen at the moment $t_k$ and is associated with the claim incurred in the interval $\Delta t_k$ and its subsequent development before this date.

It should be remembered that an insurance policy could include deductibles which essentially truncate the space of the random value $s$. Indeed, if the insurance policy provides for a deductible at the level of $d_k$, then the insurance company does not record claims $s<d_k$ in the Claim Register and, accordingly, does not pay for them since this is the liability of the policyholder. In other words, the value $s$ is seen under the random condition $s \notin T_k^s$, where the set $T_k^s=\{s:s \le d_k\}$ is called the truncation set or the truncating set (see ~\cite{Baskakov2019}) associated with its $S_k$.

However, when considering the 2D random vector $(s, \tau)$, the situation changes fundamentally. The value of the second component $\tau$ of the vector $(s, \tau)$ may be seen at any time since the set $C_k^{\tau}$, which its actual value $\tau_k$ belongs to, is always known. Therefore, ultimate (i.e. not truncated) information in form of the censoring set $C_k$ about the realization of the $S_k$ vector $(s, \tau)$ is available at any moment $t$, and the family
\begin{equation}                                                                                                          \label{equ:7}
  \{C_k\}, k=1, \dots , n.
\end{equation}
forms a censored sample from the distribution $\textbf{F} (s, \tau)$.

Thus, when considering the 2D vector $(s,\tau)$, a deductible changes the from of censoring sets $C_k$, but does not affect the sample structure. It remains censored, and in case of 1D, a deductible transforms the sample type from censored to truncated-censored.

$\textbf{Example.}$ Let us assume that an insurer organized the collection of insurance statistics in accordance with the scheme outlined and it is required to make actuarial calculations based on the data obtained on 31.06.2019. Let us assume that Table~\ref{tab:1} represents the data collected for the insurance policy No. $i$ with a deductible of 10,000.

\begin{table}
\begin{center}
\caption{\label{tab:1} Source statistics}
\begin{tabular}{lccc}
\noalign{\smallskip}\hline\noalign{\medskip}
\multicolumn{2}{l}{Indicator} & Symbol	& Value \\
\noalign{\medskip}\hline\noalign{\smallskip}
\multicolumn{2}{l}{Policy start date} & 	$t_i^1$		& January 1, 2017 \\[\smallskipamount]
\noalign{\smallskip}\hline\noalign{\smallskip}
\multicolumn{2}{l}{Policy early termination/cancellation date} & 	$t_i^2$	& January 1, 2018 \\ [\smallskipamount]

\noalign{\smallskip}\hline\noalign{\smallskip}
\multirow{2}*{Occurrence date of $j$ insurance claim} & $j=1$ & $\tau_{i1}^1$	 & May 16, 2017 \\
\noalign{\smallskip}\cline{2-4}\noalign{\smallskip}
& $j=2$	& $\tau_{i2}^1$  & November 21, 2017 \\

\noalign{\smallskip}\hline\noalign{\smallskip}
\multirow{2}*{Report date of $j$ insurance claim} & $j=1$ & $\tau_{i1}^2$	 & September 3, 2017 \\
\noalign{\smallskip}\cline{2-4}\noalign{\smallskip}
& $j=2$	& $\tau_{i2}^2$  & January 7, 2019 \\

\noalign{\medskip}\hline\noalign{\medskip}
\multirow{2}*{$j$ claim settlement date } & $j=1$ & $\tau_{i1}^3$	 & August 12, 2018 \\
\noalign{\smallskip}\cline{2-4}\noalign{\smallskip}
& $j=2$	& $\tau_{i2}^3$  & --- \\
\noalign{\smallskip}\hline\noalign{\smallskip}

\multirow{2}*{Total payments as of June 31, 2019 for $j$ claim } & $j=1$ & $s_{i1}(t)$	 & 95\\
\noalign{\smallskip}\cline{2-4}\noalign{\smallskip}
& $j=2$	& $s_{i2}(t)$ & 39 \\
\noalign{\smallskip}\hline\noalign{\smallskip}
\end{tabular}
\end{center}
\end{table}

Let us consider the procedure for censored sampling in the form (\ref{equ:7}) according to data in Table~\ref{tab:1}. For simplicity, let us assume that

(a) for this type of insurance, no more than one insurance claim may occur within a calendar month, and

(b) under the insurance contract, the limitation period $L$ is max. two years.

Based on the assumption (a), we may take $\Delta t_k$ as one month and create the sample $\{C_k \}$, $k=1,\dots, 12$, where each element corresponds to every month of the policy validity period's 12 months (see Table~\ref{tab:2}), where time is measured in whole months passed from the policy start date $\tau_k^1$).

\begin{table}
\begin{center}
\caption{\label{tab:2} Multidimensional censored vector sample ($s_k, \tau_k, \tau_k^*$)}
\begin{tabular}{cccccc}
\noalign{\smallskip}\hline\noalign{\smallskip}

$k$ &	$t_k=t-\tau_k^1$ &	$\tau_k=\tau_k^2-\tau_k^1$	& $\tau_k^*=\tau_k^3-\tau_k^1$ &	 $s_k \cdot 10^{-3}$ & $\delta_k$ \\
\noalign{\smallskip}\hline\noalign{\smallskip}
1   &	29	& $\infty$ &	$\infty$ &	0       &	0 \\
\noalign{\smallskip}\hline\noalign{\smallskip}
2	  & 28	&	$\infty$ &	$\infty$ &	0       &	0 \\
\noalign{\smallskip}\hline\noalign{\smallskip}
3	  & 27	&	$\infty$ &	$\infty$ &	0       &	0 \\
\noalign{\smallskip}\hline\noalign{\smallskip}
4	  & 26	&	$\infty$ &	$\infty$ &	0       &	0 \\
\noalign{\smallskip}\hline\noalign{\smallskip}
5	  & 25	& 4	       &  15	     & 95	      & 0 \\
\noalign{\smallskip}\hline\noalign{\smallskip}
6	  & 24	&	$\infty$ &	$\infty$ &	0       &	0 \\
\noalign{\smallskip}\hline\noalign{\smallskip}
7	  & 23	& $>23$    &	$>23$	   & $\ge 10$	& 2 \\
\noalign{\smallskip}\hline\noalign{\smallskip}
8	  & 22	& $>22$    &  $>22$	   & $\ge 10$	& 2 \\
\noalign{\smallskip}\hline\noalign{\smallskip}
9	  & 21	& $>21$    &	$>21$	   & $\ge 10$	& 2 \\
\noalign{\smallskip}\hline\noalign{\smallskip}
10	& 20	& $>20$    &  $>20$	   & $\ge 10$	& 2 \\
\noalign{\smallskip}\hline\noalign{\smallskip}
11	& 19	& 14	     &  $>19$	   & $\ge 39$	& 1 \\
\noalign{\smallskip}\hline\noalign{\smallskip}
12	& 18	& $>18$    &  $>18$	   & $\ge 10$	& 2 \\
\noalign{\smallskip}\hline\noalign{\smallskip}
\end{tabular} \\[\bigskipamount]
\end{center}
\end{table}

\begin{figure}
\center
  \includegraphics[width=1\textwidth]{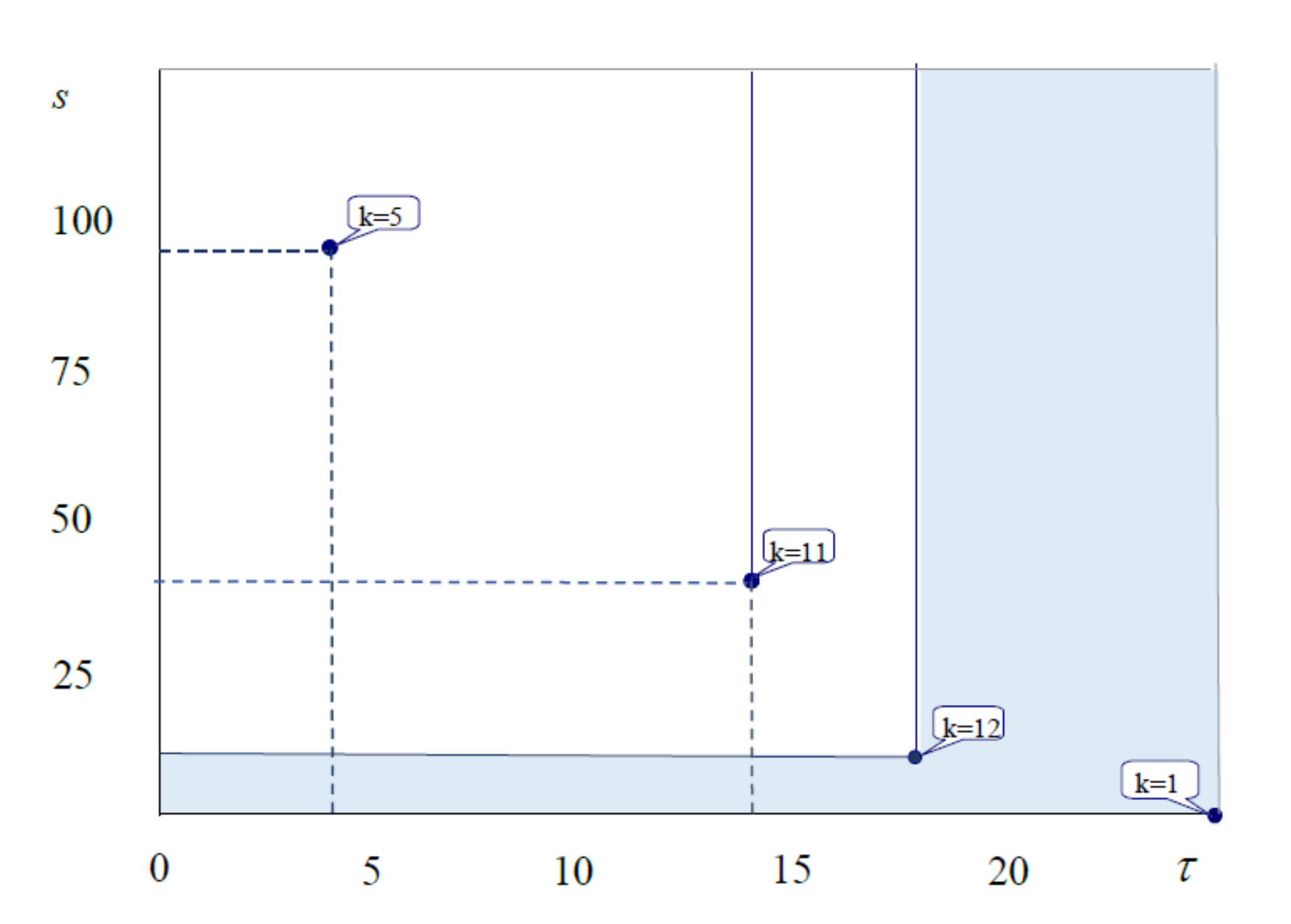}
\caption{Censoring sets of the two-dimensional vector $(s,\tau)$}
\label{fig:3}       
\end{figure}

Note that the data of Example are selected so that the sample in Table~\ref{tab:2} contains all the four possible censoring set types, e.g. corresponding to $k=1$, $5$, $11$ and $12$ (see Figure~\ref{fig:3}). The truncating set in the specified example is the same for all the elements of the sample since they all belong to the same policy with a deductible of 10,000. Thus, the expression for the censoring set may be written as

\begin{equation}                                                                                                          \label{equ:8}
C_k =\left\{
    \begin{array}{cl}
         &  (S_k,\tau_k),   \ ~~~~~~~~~~~~~~~~~~~~~~~~~~~~\mbox{if}\ \delta_k=0, \tau_k \le L, \tau_k \le t_k   \\
         &  (0, \infty),    \ ~~~~~~~~~~~~~~~~~~~~~~~~~~~~~~\mbox{if}\ \delta_k=0, \tau_k > L, t_k > L \\
         &  (s, \tau):  s \ge s_k(t), \tau=\tau_k, \ ~~~~~~~\mbox{if}\ \delta_k=1, \tau_k \le L, t_k \le t_k     \\
         &  (s, \tau): \left\{
            \begin{array} {cl}
                & s \le D_k, \tau \in (0, L], \\
                & s > D_k, \tau \in (t_k, L], \\
                & s = 0, \tau = \infty, \\
            \end{array}
         \right. \ \mbox{if}\ \delta_k=2, \tau_k > t_k, t_k < L          \\
    \end{array}
\right.
\end{equation}

\section{Estimation of the distribution function  }
\label{sec:2}

Quasi-empirical distribution~\cite{Baskakov1996} (or qED estimator~\cite{Baskakov2019})   is used to estimate the distribution $\textbf{P} (B)$ of the sample~(\ref{equ:7})
\begin{equation}                                                                                                          \label{equ:9}
\textbf{P}_n^* (B)=\frac{1}{n} \sum_{k=1}^n \textbf{P}_n^* (B|C_k ),~B \subset R^2,~\textbf{P}_n^* \in \mathcal{P}.
\end{equation}
It is determined as the solution of a functional
$$
\textbf{P}_n (B)=\frac{1}{n}  \sum_{k=1}^n \textbf{P} (B|C_k ),~B \subset R^2,~\textbf{P} \in \mathcal{P}.
$$
where $\textbf{P} (B|C_k )$ is the conditional distribution; $\mathcal{P}$ is the class of all possible distributions on $R^2$; $B$ is the measurable set, $B \subset R^2$.

The estimate $\textbf{P}_n^* (B)$ generalizes the usual empirical distribution, it is a generalized estimate of maximum likelihood $\textbf{P}(B)$ against $\mathcal{P}$ and, under certain conditions, it inherits its basic properties: consistency, asymptotic normality, etc.

To construct an estimate of the distribution function~(\ref{equ:9}), one can use an iterative procedure of the EM-algorithm type~\cite{Dempster1977}, which includes the following steps.
\begin{enumerate}
\item Set the initial approximation of the estimate $\textbf{P}_n^{(0)} (B)$, ~$B \subset R^2$, for example, uniform approximation.
\item Calculate the new value $\textbf{P}_n^{(1)} (B)$ as follows:
$$
\textbf{P}_n^**{(1)} (B)=\frac{1}{n}  \sum_{k=1}^n \textbf{P}_n^{(0)} (B|C_k ).
$$
\item Return to step 2, replacing $\textbf{P}_n^{(0)} (\cdot)$ with $\textbf{P}_n^{(1)} (\cdot)$, etc.
\item Calculations are complete with the set accuracy achieved.
\end{enumerate}

Note that if $B=\{S,\tau: S<s, \tau<t\}$, then $\textbf{P}(B)=\textbf{P}(S<s,~\tau<t)=\textbf{F}(s,t)$, i.e. is the distribution function in the conventional sense, that is, the unique, unambiguous, real, and non-negative function of the point $\{s,t\} \subset R^2$. The estimate $\textbf{F}_n^* (s,t)$ can be constructed explicitly using the above algorithm and the following formulas for calculating probabilities:
$$
\textbf{F}_n^{(0)} (s,t|C_k )=\left\{
    \begin{array}{cl}
& 1,                \ ~~~~~~~~~~~~~~~~~~~~~~~~~~~~~~~~~~~~~~~~~~~~~~~~~~~~~~~~~~~~if~~ \delta_k=0 \\\\
&\displaystyle \frac{(\textbf{F}_n^{(0) } (s,\tau_k )-\textbf{F}_n^{(0) } (s_k,\tau_k ))}{(\textbf{F}_n^{(0)} (\infty,\tau_k )-\textbf{F}_n^{(0)} (s_k,\tau_k ) )},  \ ~~~~~~~~~~~~~~~~~~~~~~~~~if~~ \delta_k=1 \\\\
&\displaystyle \frac{(\textbf{F}_n^{(0)} (s,t)-\textbf{F}_n^{(0)} (s,\tau_k )+\textbf{F}_n^{(0)} (min(s,d_k),\tau_k )  )}{(1-\textbf{F}_n^{(0)} (\infty,\tau_k )+\textbf{F}_n^{(0)} (d_k,\tau_k ) )},\ ~if~~   \delta_k=2 \\
  \end{array}
\right.
$$

if $B \cap C_k \ne \emptyset$ (otherwise $\textbf{P}_n^{(0)} (B|C_k )=0$), obtained in view of the expression~(\ref{equ:8}) for censoring sets. Marginal distributions of claims $\textbf{F}_n^* (s)$ and report dates $\textbf{F}_n^* (t)$ can be calculated as follows:
$$
\textbf{F}_n^* (s_k )=\textbf{F}_n^* (s_k,\infty)
$$
and
$$
\textbf{F}_n^* (t_k )=\textbf{F}_n^* (\infty,t_k ).
$$

The nonparametric estimation algorithm for two-dimensional distribution function $\textbf{F}_n^* (s,t)$ based on type~(\ref{equ:7}) censored data is implemented in the SAS environment using SAS/IML. Unlike the basic algorithm, it uses grouped data, so the computational complexity is almost independent of the number of insurance policies. This allows to process data of not just a single company, but also of the insurance industry as a whole.

\section{Applied tasks}
\label{sec:3}

Let us consider solutions to some applied tasks based on the proposed methodology. The real data of a number of Russian general insurance companies were used as initial statistical information. These data combine information about 271,674 same-type contracts to build a two-dimensional censored sample (see Table~\ref{tab:3}). The sample scope (exposure) is 85,244,280 days. The number of claims reported and settled ($\delta=0, \tau<\infty$) is 11,196, the number of outstanding claims ($\delta=1$) is 1,393. The number of days with claims not reported on the reporting date but that may have occurred and may be claimed in the future ($\delta=2$) is 69,693,135, and there have been no claims in the remaining 15,538,556 days ($\delta=0, \tau=\infty$) and will never be since the 3-year limitation period has expired. The total amount of claims paid at the moment of data collection is 814,218,985$\rouble$.

\begin{figure}
\center
  \includegraphics[width=1\textwidth]{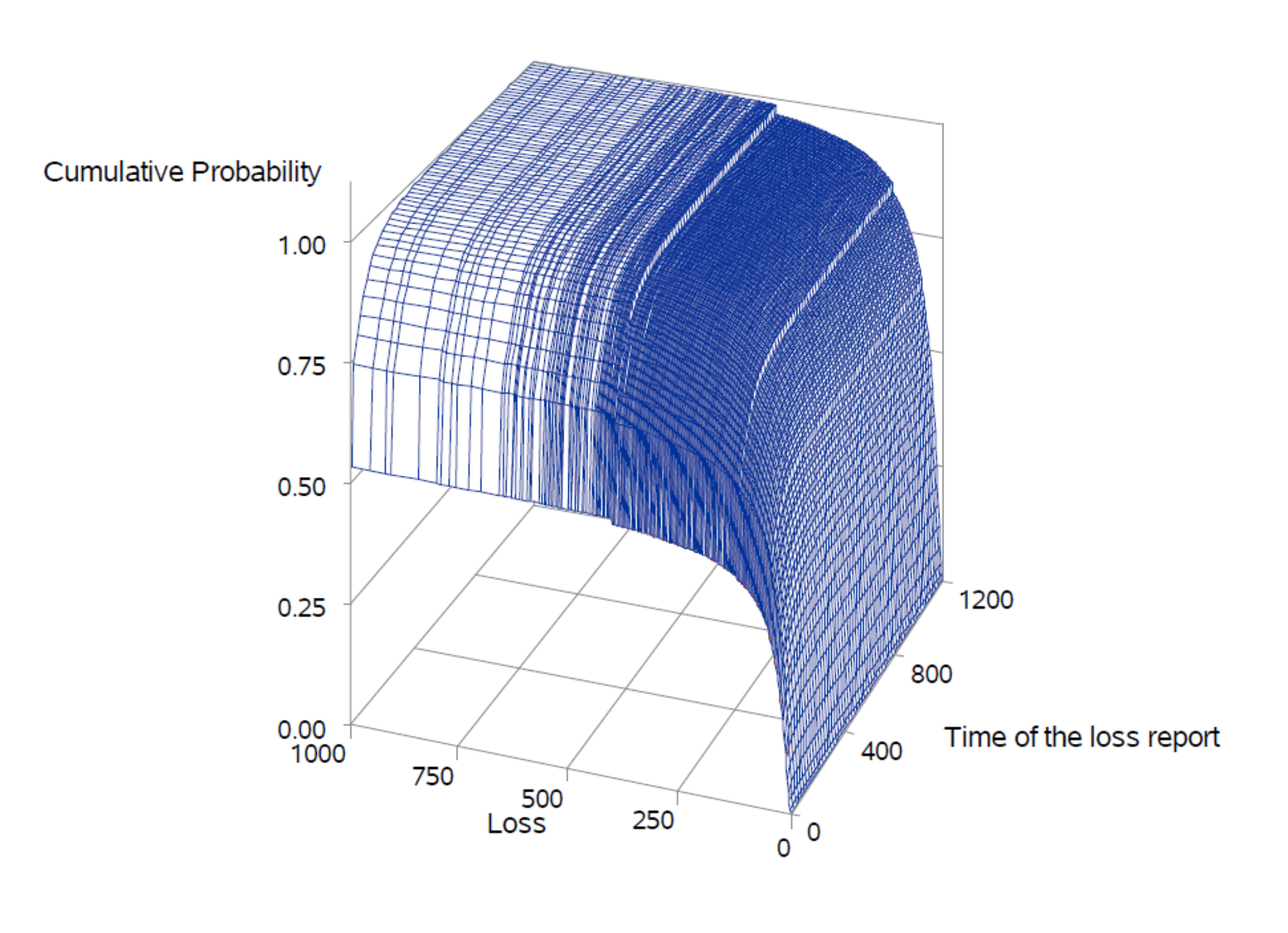}
\caption{Estimate of two-dimensional distribution function $\textbf{F}(s,\tau|\tau<\infty)$}
\label{fig:4}       
\end{figure}

\begin{figure}
\center
  \includegraphics[angle=0,width=1\textwidth]{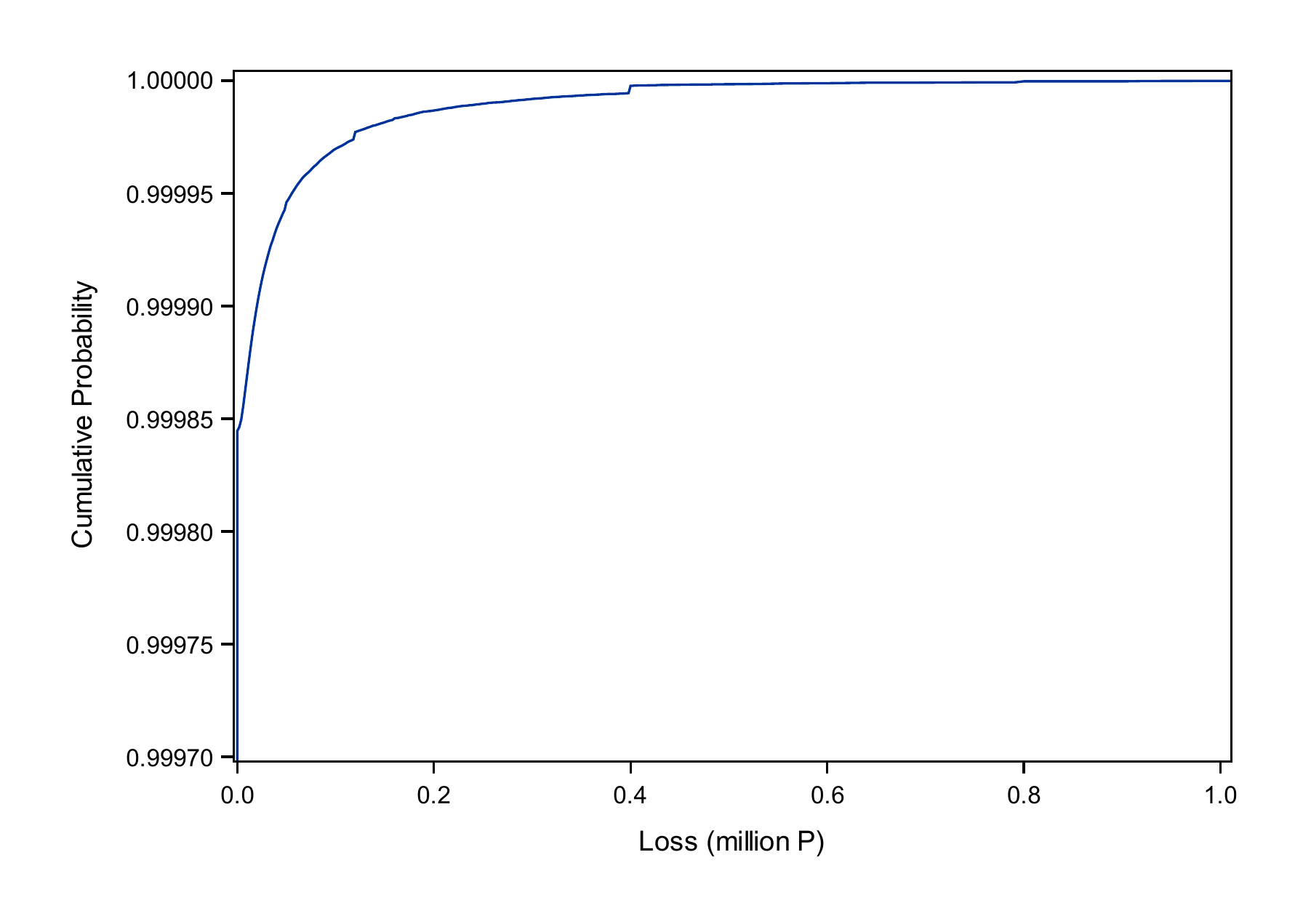}
\caption{Estimate of the marginal distribution function $\textbf{F}(s)$}
\label{fig:5}       
\end{figure}

The distribution function $\textbf{F}(s,\tau)$ is estimated by the above method. For clarity, Figure~\ref{fig:4} shows the function's nominal variant $\textbf{F}(s,\tau|\tau<\infty)$ since the graph $\textbf{F}(s,\tau)$ is undescriptive due to the insufficient value of $\textbf{P}(\tau<\infty)=0.000555$. An estimate of the marginal distribution $\textbf{F}(s)$ is shown in Figure~\ref{fig:5}.

Estimates of the distributions $\textbf{F}(s,\tau)$, $\textbf{F}(s)$ and $\textbf{F}(\tau)$ make it simple to effectively solve a range of important problems the insurance company is facing. For example, using the estimate of the claim distribution function $\textbf{F}(s)$ the formula~(\ref{equ:1}) can be utilized to calculate its expected value $\textbf{M}(s)=11.141814\rouble$ or estimate the probability $\textbf{P}(s>0)=0.000555$. These values should be interpreted
as the net insurance rate and the insurance claims frequency for an insurance period of one day. Equivalent figures for a different insurance period are calculated by multiplying them by the insurance period in days. Thus, for the 1-year insurance policy, the net premium is $11.141814 \times 365 \approx 4,067\rouble$, and the frequency is 5.68\%. The average claim amount in terms of one insurance claim is $11.141814 / 0.000555 = 71,652\rouble$. Apparently, the amount is independent of the insurance period.

\begin{small}
\begin{landscape}
\begin{longtable}{ccllllllllll}
\caption{\label{tab:3}  Two-dimensional censored vector $(s_k, \tau_k)$ sample} \\
\noalign{\medskip}\hline\noalign{\medskip}
\multirow{2}*{$s$ } & \multirow{2}*{$\delta$} & \multicolumn{10}{l}{Days from the date of insurance claim, $\tau$} \\
\noalign{\smallskip}\cline{3-12}\noalign{\smallskip}
& & 0	& 150~~~~~~~ &	300~~~~~~~ &	450~~~~~~~ &	600~~~~~~~	& 750~~~~~~~	& 900~~~~~~~ &	 1050~~~~~~ &	1200~~~~~~ & $\infty$ \\[2pt]
\endfirsthead
\caption{\label{tab:3} (continued)}\\
\noalign{\medskip}\hline\noalign{\medskip}
\multirow{2}*{$s$ } & \multirow{2}*{$\delta$} & \multicolumn{10}{l}{Days from the date of insurance claim, $\tau$} \\
\noalign{\smallskip}\cline{3-12}\noalign{\medskip}
& & 0	& 150~~~~~~~ &	300~~~~~~~ &	450~~~~~~~ &	600~~~~~~~	& 750~~~~~~~	& 900~~~~~~~ &	 1050~~~~~~ &	1200~~~~~~ & $\infty$~~~~~~~~ \\[2pt]
\endhead
\endfoot
\noalign{\smallskip}\hline\noalign{\medskip}
\endlastfoot

\noalign{\smallskip}\hline\noalign{\smallskip}
\multirow{3}*{0} & 0	& 4357	& 527 &	92 &	16 &	7	& 5	& 3 &	. &	1 &	15538556 \\
\noalign{\smallskip}\cline{2-12}\noalign{\smallskip}
& 1	& 431 &	27 &	5 &	1 &	. &	1 &	. &	. &	 . &	. \\
\noalign{\smallskip}\cline{2-12}\noalign{\smallskip}
& 2 &	4161344 &	8388417 &	8532966 &	8703561 &	8804411 &	8814772 &	8807326 &	8924637 &	4555701 & . \\ \noalign{\smallskip}\hline\noalign{\smallskip}

\multirow{2}*{50} & 0	& 3239 &	470 &	73 &	16 &	3 &	10 &	1 &	3 &	. &	. \\
\noalign{\smallskip}\cline{2-12}\noalign{\smallskip}
& 1	& 414 &	61 &	10 &	1 &	2 &	3 &	1 &	. &	. &	. \\ \noalign{\smallskip}\hline\noalign{\smallskip}

\multirow{2}*{100} & 0 &	949 &	227 &	28 &	12 &	4 &	1 &	. &	1 &	1 &	. \\
\noalign{\smallskip}\cline{2-12}\noalign{\smallskip}
& 1	& 132 &	30 &	1	& 1	& 1 &	. &	. &	1 &	. &	. \\
\noalign{\smallskip}\hline\noalign{\smallskip}

\multirow{2}*{150} & 	0	& 342 &	63 &	12 &	3 &	3 &	1 &	. &	. &	. &	. \\
\noalign{\smallskip}\cline{2-12}\noalign{\smallskip}
& 1 &	73 &	6 &	3 &	. &	. &	. &	. &	. &	. &	. \\
\noalign{\smallskip}\hline\noalign{\smallskip}

\multirow{2}*{200} & 0 & 183 &	29 &	6 &	. &	. &	. &	. &	. &	. &	. \\
\noalign{\smallskip}\cline{2-12}\noalign{\smallskip}
& 1	& 43 &	8 &	3 &	. &	. &	. &	. &	. & .	 & . \\
\noalign{\smallskip}\hline\noalign{\smallskip}

\multirow{2}*{250} & 0 & 103	& 18 &	4 &	1 &	. &	. &	. &	. &	. &	. \\
\noalign{\smallskip}\cline{2-12}\noalign{\smallskip}
& 1	& 32	 & 3	 & 1	 &	. &	. &	. &	. &	. & .	 & . \\
\noalign{\smallskip}\hline\noalign{\smallskip}

\multirow{2}*{300} & 0 & 95	& 9	& .	& 1	 &	. &	. &	. &	. &	. &	. \\
\noalign{\smallskip}\cline{2-12}\noalign{\smallskip}
& 1	& 17	& 7 &	1 &	. &	. &	. &	. &	. & .	 & . \\
\noalign{\smallskip}\hline\noalign{\smallskip}

\multirow{2}*{350} & 0 & 	55 &	6 &	1 &	2 &	. &	. &	. &	. &	. &	. \\
\noalign{\smallskip}\cline{2-12}\noalign{\smallskip}
& 1	& 13 &	. &	. &	. &	. &	. &	. &	. & .	 & . \\
\noalign{\smallskip}\hline\noalign{\smallskip}

\multirow{2}*{400} & 0 & 141	& 26 &	4 &	. &	. &	. &	. &	. &	. &	. \\
\noalign{\smallskip}\cline{2-12}\noalign{\smallskip}
& 1	& 22	& 2	& 3 &	. &	. &	. &	. &	. & .	 & . \\
\noalign{\smallskip}\hline\noalign{\smallskip}

\multirow{2}*{450} & 0 & 11 &	1 &	. &	. &	. &	. &	. &	. &	. &	. \\
\noalign{\smallskip}\cline{2-12}\noalign{\smallskip}
& 1	& 10 &	. &	. &	. &	. &	. &	. &	. & .	 & . \\
\noalign{\smallskip}\hline\noalign{\smallskip}

\multirow{2}*{500} & 0 & 5 &	1 &	. &	. &	. &	1 &	. &	. &	. &	. \\
\noalign{\smallskip}\cline{2-12}\noalign{\smallskip}
& 1	& 7 &	1 &	. &	. &	. &	. &	. &	. & .	 & . \\
\noalign{\smallskip}\hline\noalign{\smallskip}

\multirow{2}*{650} & 0 & 22 &	1 &	. &	. &	. &	. &	. &	. &	. &	. \\
\noalign{\smallskip}\cline{2-12}\noalign{\smallskip}
& 1	& 15 &	. &	. &	. &	. &	. &	. &	. & .	 & . \\

\end{longtable}
\end{landscape}
\end{small}

$\textbf{Claims reserves.}$ Next, let us consider the problem of estimating claims reserves including the outstanding claims reserves and incurred but not reported reserves (IBNR). Note that the calculations already performed are sufficient to estimate the claims reserves (liabilities outstanding at the reporting date). Indeed, knowing the average claim amount $\textbf{M}(s)$, the exposure in days and the amount of claims paid at the reporting date, the claims reserve is estimated as follows:
\begin{equation}                                                                                                          \label{equ:10}
11.141814 \times 85,244,280-814,218,985=135,556,927\rouble.	
\end{equation}

For comparison, alternative calculations have been performed using the most popular methods of claims reserving~\cite{Benjamin1987} including the chain-ladder method, the Bornhuetter-Ferguson method \cite{Schmidt2008} and the Frequency\&Severity method. These methods are based on the cumulative paid claim development triangle (see Table~\ref{tab:4}). The bottom line of this table adds the development factors estimates obtained by the chain-ladder method.

\begin{table}
\begin{center}
\caption{\label{tab:4} Cumulative paid claim development triangle, in million rouble}
\begin{tabular}{c|ccccccccc}
\noalign{\smallskip}\hline\noalign{\smallskip}
	& 6 &	12 &	18 &	24 &	30 &	36 &	42 &	48 &	54 \\
\noalign{\smallskip}\hline\noalign{\smallskip}
6	& 4.70	& 14.33	& 17.25	& 18.64	& 19.66	& 20.15	& 20.48	& 20.85	& 20.95 \\
12 &	22.52	& 49.23	& 55.78	& 59.33	& 60.49	& 61.70	& 63.40	& 63.83 & \\
18 &	44.44	& 73.67	& 80.15	& 83.45	& 85.24	& 87.24	& 88.64		& 	&  \\
24 &	52.81	& 88.55	& 98.61	& 103.41	& 106.43	& 107.36	& 	& 	& \\		
30 &	56.49	& 92.36	& 99.79	& 105.32	& 106.33	& 	& 	& 	& 	\\		
36 &	70.28	& 117.76	& 133.06	& 137.58	& 	& 	& 	& 	& 	\\		
42 &	71.52	& 117.21	& 125.47	& 	&  	& 	& 	& 	& 		\\			
48 &	65.74	& 107.68	& 	&	  &  	& 	& 	& 	& 	\\					
54 &	56.42	& 	& 	& 	& 	& 	& 	& 	& \\
\noalign{\smallskip}\hline\noalign{\smallskip}					
\noalign{\smallskip}\hline\noalign{\smallskip}
		& 1.701	& 1.103	& 1.048	& 1.022	& 1.017	& 1.020	& 1.010	& 1.005	& 1.000 \\
\noalign{\smallskip}\hline\noalign{\smallskip}

\end{tabular} \\[\bigskipamount]
\end{center}
\end{table}

Applying the indicated reserving methods to the data in Table~\ref{tab:4}, the point estimate (best estimate) of the claims reserves amount is obtained, which is 126 million $\rouble$, and the range of reasonable estimates of the claims reserves is from 116.7 to 147.6 million $\rouble$. Note that the claims reserves estimate~(\ref{equ:10}) which was previously obtained on the basis of the methodology proposed in this paper falls within a range of reasonable estimates and slightly differs from the estimate of the claims reserves.

$\textbf{IBNR and outstanding claims reserves.}$ Let us assume that we know the two-dimensional distribution function $F(s,\tau)$, where $s$ is the amount of the insurance claim and $\tau$ is the interval between the claim occurrence and report dates. The function may be estimated using the data of the insurance portfolio being examined - like it was done above, - or it may be obtained a priori by generalizing the results of previous calculations based on similar data, or otherwise (the method is not relevant for the case). Let us have the censored sample~(\ref{equ:7})  e.g. in form of Table~\ref{tab:2}.

Let us consider the day of the insurance claim occurrence $\tau_k^1$. If at the date t the insurance claim occurred at the moment $\tau_k^1$ is not reported, then the date of the report $\tau>t_k$ and, in particular, $\tau=\infty$, if the insurance claim did not occur.

The distribution function of the claim occurred at the moment $\tau_{ik}^1$ and reported in the interval $[t_1, t_2 )$, provided that it was not announced as of the reporting date $t$, is
\begin{equation}                                                                                                          \label{equ:11}
\textbf{F}_k (s)=1-\frac{(\textbf{F}(\infty,t_{2k} )-\textbf{F}(\infty,t_{1k} ))}{(1-\textbf{F}(\infty,t_k ) )}+\frac{(\textbf{F}(s,t_{2k} )-\textbf{F}(s,t_{1k} ))}{(1-\textbf{F}(\infty,t_k ) )}.
\end{equation}

The first addends not dependent on the claim amount correspond to the probability that the claim in the interval $[t_1, t_2 )$ will not be reported, and the last addend is equal to the probability that the claim reported in the said interval is less than or equal to $s$. Note that for the interval $[t,\infty)$, the claim distribution functions are as follows
\begin{equation}                                                                                                          \label{equ:12}
\textbf{F}_k (s)=\frac{(\textbf{F}(s,\infty)-\textbf{F}(s,t_k ))} {(1-\textbf{F}(\infty,t_k ) )}.                                                                               \end{equation}

The expected claim reported in the interval $[t_1, t_2 )$ is
\begin{equation}                                                                                                          \label{equ:13}
\textbf{M}_k (s)=\int\limits_{d_k}^\infty s \cdot d \textbf{F} (s).                                                                                       \end{equation}

The claim variance is
\begin{equation}                                                                                                          \label{equ:14}
\textbf{D}_k (s)=\int\limits_{d_k}^\infty s^2 \cdot d \textbf{F} (s)-\textbf{M}_k^2 (s).                                                                    \end{equation}

The probability of the claim $s>d_k$ being reported in the interval $[t_1, t_2 )$ is
\begin{equation}                                                                                                          \label{equ:15}
\textbf{P}_k (s)=\frac{(\textbf{F}(\infty,t_{2k} )-\textbf{F}(\infty,t_{1k} )-\textbf{F}(d_k,t_{2k} )+\textbf{F}(d_k,t_{1k}))}{(1-\textbf{F}(\infty,t_k ) )}.
\end{equation}

Similar reasoning for other days with the insurance claim date $\tau_k^1, k=1, \dots, n$ allows for calculating the expected claim by single policy or by the whole portfolio.
\begin{equation}                                                                                                          \label{equ:16}
\textbf{M}_t (s)=\sum_{k=1}^n \textbf{M}_k (s)                                                                                                \end{equation}
and its variance
\begin{equation}                                                                                                          \label{equ:17}
\textbf{D}_t (s)=\sum_{k=1}^n \textbf{D}_k (s),                                                                                               \end{equation}
as well as the expected number of claims
\begin{equation}                                                                                                          \label{equ:18}
N_t=\sum_{k=1}^n \textbf{P}_k (s).                                                                                                     \end{equation}

Here, the index t means that statistics~(\ref{equ:16})~-~(\ref{equ:18}) are calculated on the condition that the claim was not reported at the reporting date $t$.

Let us consider the definition of the IBNR in form~(\ref{equ:1}). It is equivalent to expression~(\ref{equ:16}), where the expected loss (expected value of insurance payments) is used as addends $\textbf{M}_k (s), k=1, \dots, n$ in the interval $[t,\infty)$, including the costs of settling the claims occurred on reporting date $t$ or in periods preceding it, but not reported in the prescribed manner to the insurer.

Note that the qED estimator~(\ref{equ:9}) of the function $F(s,\tau)$, as the claim expected value~(\ref{equ:16}), its variance~(\ref{equ:17}) and the number of claims incurred but not reported~(\ref{equ:18}) are all additive estimates, in the sense that they are calculated as the sum of similar estimates built separately for each sample element. The additivity allows using the proposed methodology for solving problems connected with the redistribution of reserves between individual segments of insurance portfolio and/or time intervals. To do this, it is necessary to make calculations using formulas~(\ref{equ:11})~-~(\ref{equ:15}) for the relevant segment of the insurance portfolio and/or time interval and to sum up the results.

In particular, if we consider a system of non-overlapping time intervals $[t_i, t_{i+1} )$, $i=1, 2, 3, \dots$ such that
$$
\bigcup\limits_{i=1}^\infty [t_i, t_{i+1})=[t,\infty),
$$
IBNR may be redistributed by these intervals depending on the expected number of claims incurred but not reported, which will be reported in certain periods, e.g. in each subsequent quarter after the reporting date $t$.

Similar comments apply to formulas~(\ref{equ:17}) and (\ref{equ:18}) which can be used to estimate the IBNR variance and the number of claims reported in any predetermined time interval $[t_1, t_2)$ after the reporting date $t$.

Let us consider the problem of building an interval estimate of the reserve. In theory, the solution is quite simple: using the generating functions and the reserve distribution function~(\ref{equ:13}) for a single day, the formula is obtained for calculating the reserve distribution function for the entire insurance portfolio and for determining tolerance range. However, this approach is not suitable for actual calculations due to the computational complexity. Therefore, to build the interval estimate of the reserve, we assume that its distribution is Gaussian (the grounds for this will be shown later), and the lower $s_L$ and upper $s_U$ tolerance range equal, respectively, to
\begin{equation}                                                                                                          \label{equ:19}
s_L=\textbf{M}_t (s)-u_{\frac{(1-p)}{2}} \sqrt{\textbf{D}_t (s) }~~ and~~s_U=\textbf{M}_t (s)+u_{\frac{(1-p)}{2}} \sqrt{\textbf{D}_t (s) },
\end{equation}
where $u_\alpha$ is the quantile of the standard normal distribution of level $\alpha$.

Calculation of IBNR and the related indicators was done according to the above method. Then, for $\alpha=0.95$, we have: $\textbf{M}_t (s)=35,759,781\rouble \pm 4,057,684\rouble$, $\textbf{D}_t (s) )=2,466,898\rouble$ and $N_t=448.21$. The estimated average amount of claim incurred but not reported is 79,785  and the average amount of all expected claims is 71,652$\rouble$.

\begin{figure}
\center
  \includegraphics[angle=0, width=1\textwidth]{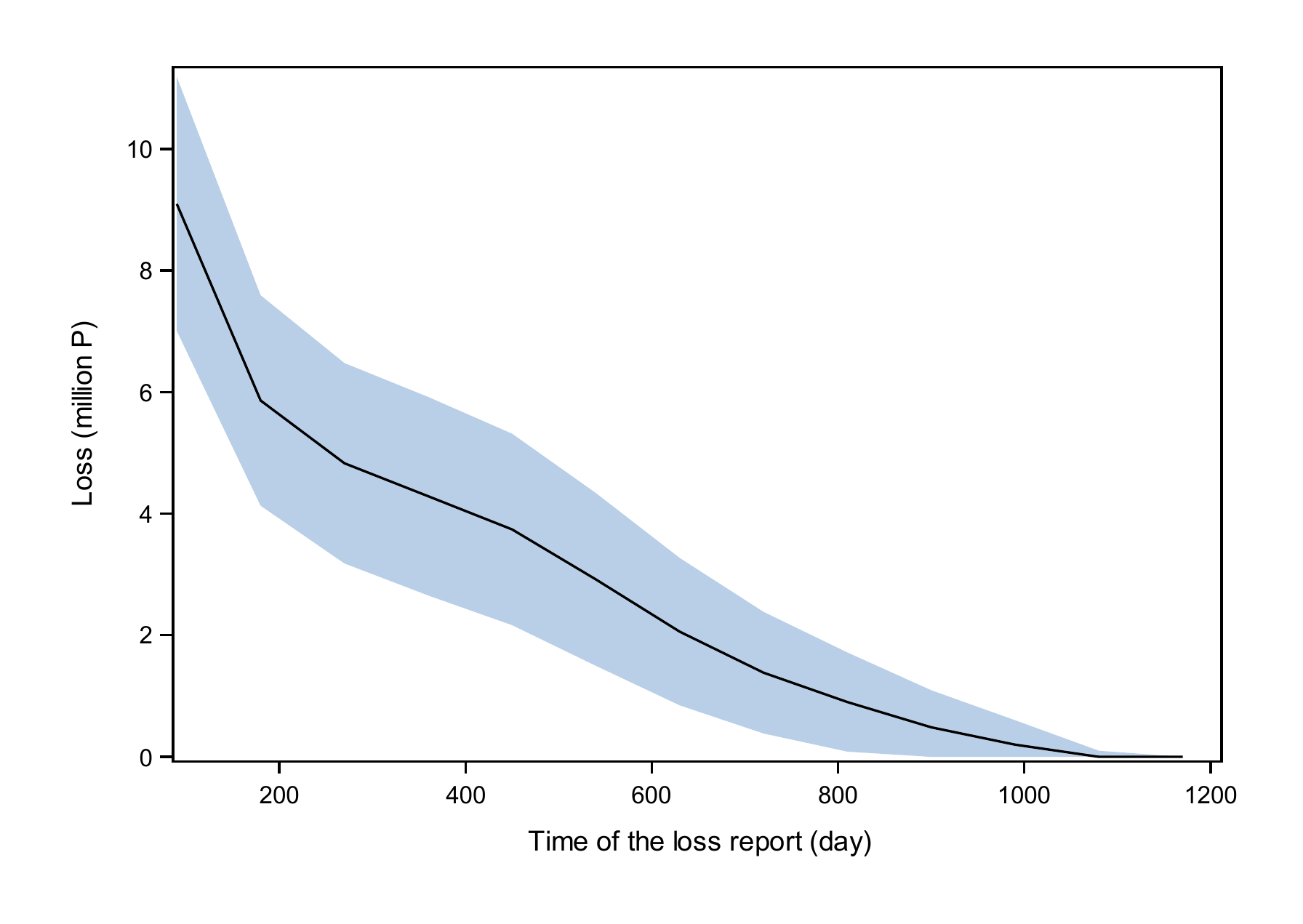}
\caption{Distribution of expected claims over time
(The mean of IBNR and 95\% confidence bands)}
\label{fig:6}       
\end{figure}

The distribution of claims over time is shown in Table~\ref{tab:5}. Figure~\ref{fig:6} shows the point and interval estimate of the reserve necessary to cover the claims that will be reported in the corresponding time interval after the reporting date.

\begin{table}
\begin{center}
\caption{\label{tab:5} IBNR distribution over time}
\begin{tabular}{*{6} c }
\noalign{\smallskip}\hline\noalign{\smallskip}

\multirow{2}* {Time}  & \multicolumn{3}{c}{Expected claim in million \rouble} & \multirow{2}* {Standard}   & \multirow{2}* {Expected}  \\
\noalign{\smallskip}\cline{2-4}\noalign{\smallskip}
interval in  & Lower  & Average  & Upper  & deviation in   & number of \\
days  &  limit &  value &  limit & million \rouble & claims\\
\noalign{\smallskip}\hline\noalign{\smallskip}

[0, 90)    & 7.006	& 9.095	& 11.184& 1.270	& 143.91\\
\noalign{\smallskip}\hline\noalign{\smallskip}
[90, 180)	 & 4.131	& 5.861	& 7.592	& 1.052 & 83.25 \\
\noalign{\smallskip}\hline\noalign{\smallskip}
[180, 270) & 3.181  & 4.829 & 6.477 & 1.002 & 62.41 \\
\noalign{\smallskip}\hline\noalign{\smallskip}
[270, 360) & 2.654  & 4.286 & 5.919 & 0.992 & 50.21 \\
\noalign{\smallskip}\hline\noalign{\smallskip}
[360, 450) & 2.166  & 3.741 & 5.315 & 0.957 & 39.65 \\
\noalign{\smallskip}\hline\noalign{\smallskip}
[450, 540) & 1.497  & 2.918 & 4.338 & 0.864 & 28.29 \\
\noalign{\smallskip}\hline\noalign{\smallskip}
[540, 630) & 0.847  & 2.059 & 3.271 & 0.737 & 18.18 \\
\noalign{\smallskip}\hline\noalign{\smallskip}
[630, 720) & 0.384  & 1.384 & 2.384 & 0.608 & 11.22 \\
\noalign{\smallskip}\hline\noalign{\smallskip}
[720, 810) & 0.085  & 0.901 & 1.716 & 0.496 & 6.73  \\
\noalign{\smallskip}\hline\noalign{\smallskip}
[810, 900) & 0.000  & 0.486 & 1.095 & 0.371 & 3.25  \\
\noalign{\smallskip}\hline\noalign{\smallskip}
[900, 990) & 0.000  & 0.200 & 0.602 & 0.244 & 1.11  \\
\noalign{\smallskip}\hline\noalign{\smallskip}
[990, 1080)& 0.000  & 0.000 & 0.000 & 0.000 & 0.00  \\
\noalign{\smallskip}\hline\noalign{\smallskip}

\end{tabular}
\end{center}
\end{table}

Incurred outstanding claims can be estimated in two ways, firstly, as the difference between the claims reserve and IBNR, i.e.
$$
135,556,927-35,759,781=99,797,146\rouble.
$$
And, secondly, as the sum of expected outstanding claims, provided that $S>s(t_k)$, that is,
$$
\textbf{M}_t (s)=\sum_{k=1}^n \textbf{M}_k (s|S>s(t_k ),\delta_k=1).
$$
The second calculation option theoretically allows us to estimate the distribution function of the reported outstanding claims and the interval estimate of the corresponding reserve.

\section{Estimation accuracy}
\label{sec:4}

As regards the practical application prospects of the proposed algorithm for modeling financial flows of insurance companies, the estimation accuracy of their parameters from samples with sizes adequate to the data available to insurers is of special interest. IBNR is selected as a parameter to integrally describe the estimation accuracy of financial flows. In fact, to calculate the parameter it is necessary to estimate the two-dimensional distribution function $\textbf{F}(s,\tau)$, as well as some additional parameters. Therefore, the IBNR estimation accuracy allows us to assess the accuracy of all the parameters involved in its calculation. The accuracy study of the reserves estimation is carried out using stochastic modeling based on a mathematical model which imitates in sufficient detail the processes in real insurance companies, including the conclusion and termination of insurance policies, occurrence, reporting and settlement of claims.

The stochastic modelling is performed in the SAS environment under the assumption that the claim amount is log-normal, and the time between the claim occurrence and report dates is a gamma distribution (see Table~\ref{tab:6}). The insurance portfolio size ranges from 100 to 100,000 same-type policies, and the claims frequency varies from 0.03 to 0.3. For every fixed set of parameter values of the model, at least $K=1000$ samples were generated to estimate on their base the percentage error and confidence limits of IBNR estimates.

The modeling algorithm is as follows:
\begin{enumerate}
\item[i] Set the input parameters of the model, including the average claim $\textbf{M}(s)$ and claim frequency $\textbf{P}(s>0)$

\item[ii] Model the insurance portfolio of $n$ policies and calculate the amount of claims modeled as incurred but not reported

\item[iii] Form a multidimensional censored sample and build the estimate of the distribution function $\textbf{F}(s,\tau)$ according to the above method

\item[iv] The IBNR is evaluated by formula~(\ref{equ:16})

\item[v]  Repeat steps (ii) - (iv) at least $K$ times

\item[vi] Evaluate the percentage error of IBNR estimation and the corresponding confidence limits.
\end{enumerate}

\begin{table}
\begin{center}
\caption{\label{tab:6} Main parameters of the model}
\begin{tabular}{cccc}
\noalign{\smallskip}\hline\noalign{\smallskip}

\multirow{2}*{Average claim size} & \multirow{2}*{Claim frequency} &
\multicolumn{2}{c}{Distribution parameters:} \\
\noalign{\smallskip}\cline{3-4}\noalign{\smallskip}
& & {$\tau \sim \Gamma(k, \theta)$} & {$s \sim ln N(a, \sigma^2)$} \\
\noalign{\smallskip}\hline\noalign{\smallskip}
\multirow{2}*{$\textbf{M}(s)=100$}   & \multirow{2}*{$\textbf{P}(s>0)=0.2$}	 & $\theta=1000 $	 & $\sigma=0.9 $   \\
& & $ k=1$	 & $ a=lnM(s)-\frac{\sigma^2}{2}$   \\
\noalign{\smallskip}\hline\noalign{\smallskip}

\end{tabular}
\end{center}
\end{table}

Note that the sum of not reported claims calculated in step (ii) is random since it is formed up in the process of modeling as a result of a correlation of a number of random variables, theoretically it is equal to IBNR. Let us set its expected value as $\textbf{M}_t (R)$. Then the percentage error of IBNR estimation can be calculated, e.g. by formula
\begin{equation}                                                                                                          \label{equ:20}
K=\frac{\textbf{M}_t (s)}{\textbf{M}_t (R)},                                                                                                            \end{equation}
where	$\textbf{M}_t (s)$ is estimated in step (iv) by formula~(\ref{equ:16}).

\begin{figure}
\center
  \includegraphics[angle=0, width=1\textwidth]{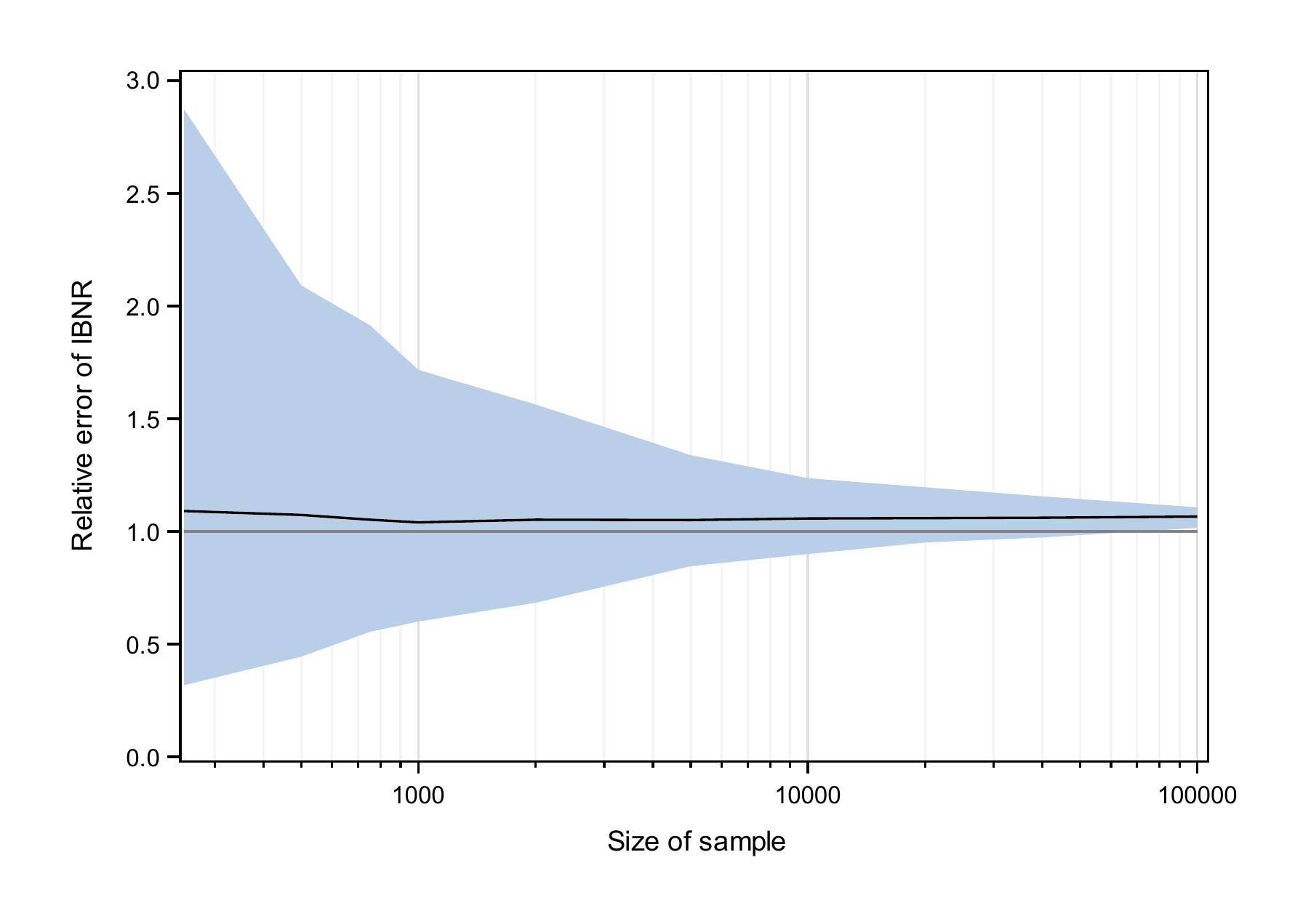}
\caption{IBNR percentage error depending on sample size}
\label{fig:7}       
\end{figure}

Figure~\ref{fig:7} shows the dependence of the relative error of IBNR estimation~(\ref{equ:20}) and the limits of its 98\% tolerance interval depending on the sample size determined by the number of policies. The graph shows that at the sample sizes taken for the modeling, the expected value of the studied parameter has a slight shift, not exceeding 5\% in absolute value. Moreover, for the samples with 500 policies and more, the proposed algorithm gives a little exaggerated IBNR estimate which can be interpreted as an actuarial reserve. The variance of the percentage error of the IBNR estimation decreases sharply with an increase of the sample size. Quantitatively, it looks as follows: with 1,000-times increase in the sample size (from 100 to 100,000 policies), the error variance decreases by more than 700 times.

\begin{figure}
\center
\includegraphics[width=1\textwidth]{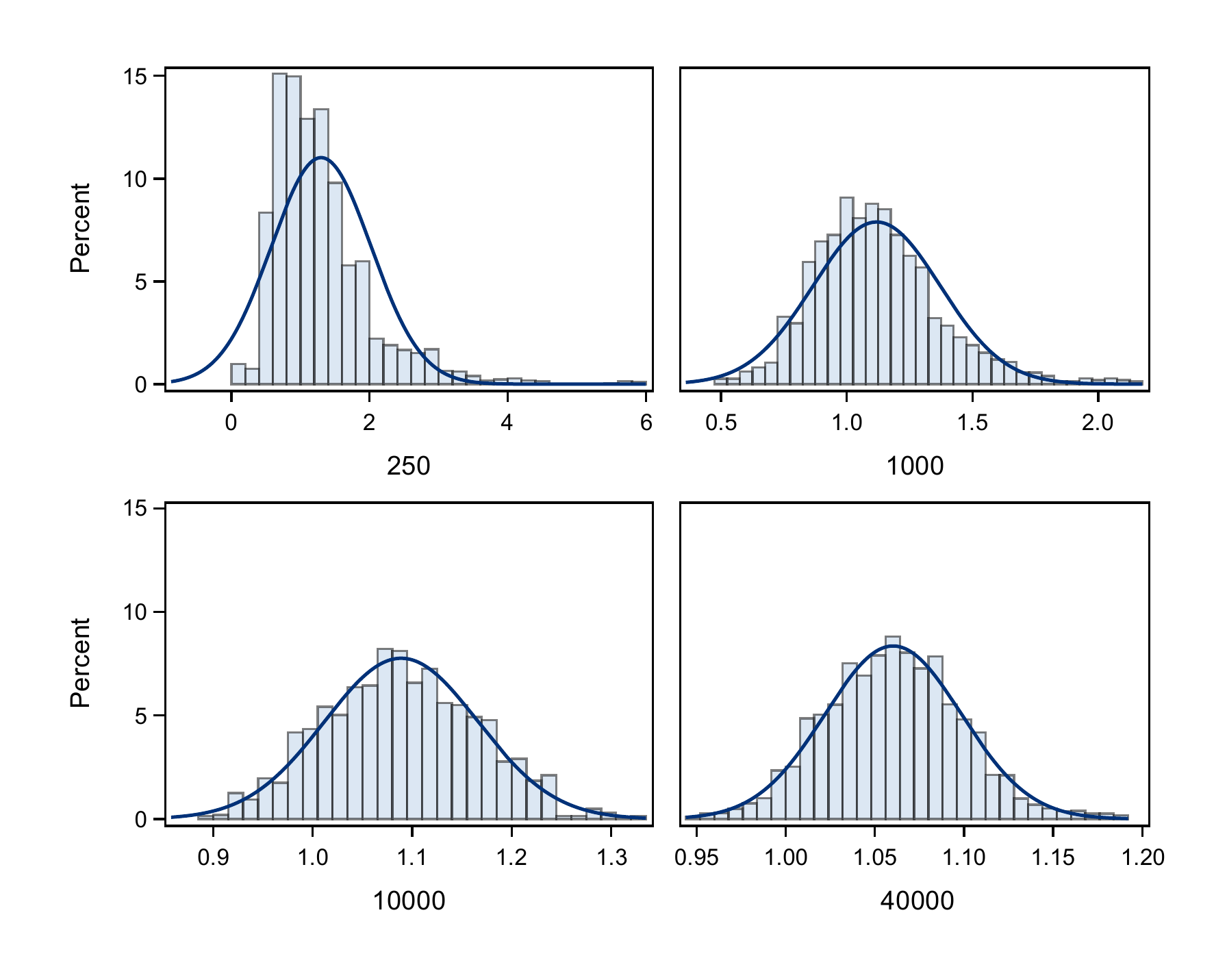}
\caption{Distribution density of IBNR normalized value depending on the sample size}
\label{fig:8}       
\end{figure}

Next, the graphs in Figure~\ref{fig:7} show that at a sample size of 2000 policies and more, the tolerance range becomes symmetrical about their expected value which almost coincides with the median in this case. The 3 sample size graphs in Figure~\ref{fig:8} clearly show the connection of the shape of density distribution of the relative error with the sample size: 250, 1000 and 10,000 policies.

Thus, the modeling results show that the IBNR estimate distribution is close to normal at sufficiently large sample sizes. This confirms that the expression~(\ref{equ:19}) can feasibly be used to calculate the tolerance limits. For example, a direct comparison of the calculation results of 98\% tolerance intervals by the formula~(\ref{equ:19}) and those obtained in the simulation (the reserve was estimated from a sample of 10,000 policies, the tolerance intervals were estimated from a sample of 1000 reserve estimates) show that the latter intervals are 5.9\% wider, and with a sample size of 40,000 policies, this difference is only 0.8\%.

\section{Summary and conclusions}
\label{sec:5}

The paper proposes an original methodology for constructing quantitative statistical models based on multidimensional distribution functions $\textbf{F}(s,\tau)$ constructed on the basis of the insurance companies' data on inshurance policies (including policies with deductible) and claims incurred. Real data of some Russian insurance companies on non-life insurance contracts illustrate some opportunities of the proposed approach. The point and interval estimates of net premium, claims frequency, claims reserves including IBNR and OCR, are thus obtained. The resulting estimate of claims reserves falls in the range of reasonable estimates calculated on the basis of traditional reserving methods (the chain-ladder method, the frequency-severity method and the Bornhuetter-Ferguson method).


The proposed methodology is based on additive estimates of a company's multidimensional distribution functions and financial indicators, in the sense that they are calculated as a sum of estimates built separately for each element of the sample (claim). This allows using the proposed methodology to model insurance companies' financial flows and, in particular, to solve the problems of reserve redistribution between particular segments of insurance portfolio and/or time intervals.

The accuracy of insurance companies' financial parameters estimate based on the proposed methods was tested by statistical modeling. IBNR was used as the test parameter. The modeling results showed a satisfactory accuracy of the proposed reserve estimates.

The advantages of the proposed methodology in comparison with traditional approaches are obvious - it makes it possible to build estimates for a greater number of a company's financial indicators with the only assumption of randomness and homogeneousness of the sample used to estimate the distribution function $\textbf{F}(s,\tau)$ (this is the most used assumption when applying most statistical methods).

\begin{figure}
\center
  \includegraphics[width=1\textwidth]{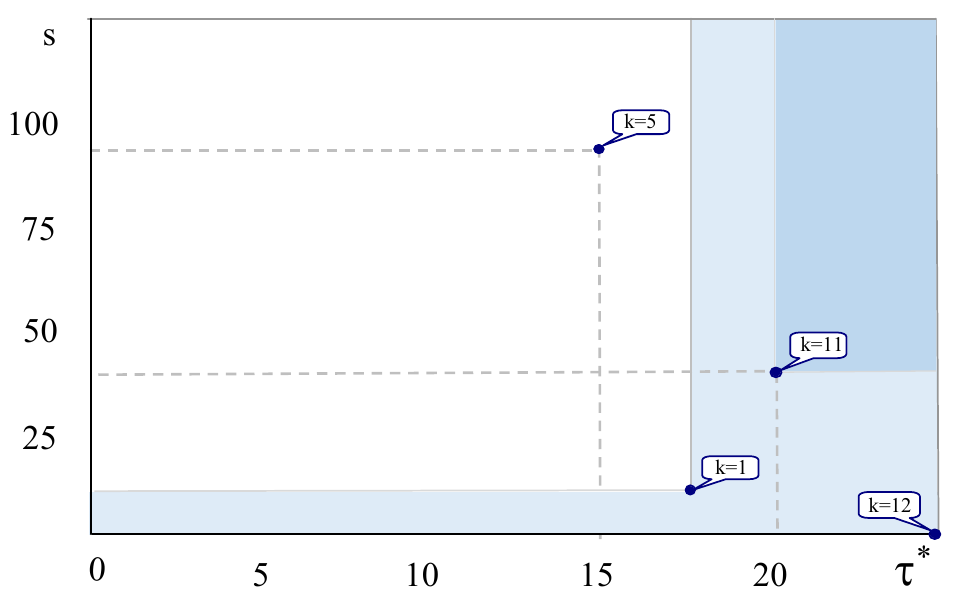}
\caption{Censoring sets of the two-dimensional vector $(s, \tau^*)$}
\label{fig:9}       
\end{figure}

Equally, the assumption of randomness and homogeneousness is the main disadvantage of the methodology since this assumption does not always hold true in practice: (1) due to inflation, the claim size depends on the calendar time and (2) when building the distribution estimate $\textbf{F}(s,\tau)$, the time of claim settlement is not considered. However claim size often correlates with settlement time. The first issue can be easily overcome by standard methods, by bringing all payments to a fixed date. The second problem can be solved only partially, e.g. by using the distribution function $\textbf{F}(s,\tau^*)$, in which the coordinate $\tau$ is replaced with the settlement period $\tau_k^*=\tau_k^3-\tau_k^1$ (see Table~\ref{tab:2}). The structure of censoring sets will be changing simultaneously since the information on reported outstanding claims becomes less certain, which objectively reduces the accuracy of the distribution function estimation. This is shown in Figure~\ref{fig:3} and Figure~\ref{fig:9} which represent the sample elements with identical values $k$.

 For one-column wide figures use

Despite these drawbacks, the proposed method is fairly resistant to certain deviations of the model from the theoretical one (in particular, to the existence of the correlation between the claim size and settlement date), and therefore can be recommended for use in actuarial practice.

The article considers only one method for processing censored samples, i.e. the qED estimator~(\ref{equ:9}). In fact, the set of methods for estimating distributions using censored data is as diverse as that using full data. It includes the parametric maximum likelihood method, the generalized minimum distance method, as well as additive, kernel, and other estimates.

The properties of statistical evaluations are largely determined by the structure and size of the censored sample, therefore it is impossible to provide one optimal evaluation method for all cases; for this reason a lot of creativity remains to be needed to solve the discussed problems.





\begin{thebibliography}{}
%


\bibitem{Baskakov2019}
Baskakov V, Bartunova A (2019) Nonparametric estimation of multivariate distribution function for truncated and censored lifetime data. Eur. Actuar. J. 9:209--239
\bibitem{Baskakova2014}
Baskakova A, Baskakov V (2014) IBRN reserves estimate on the bases of multivariate censored
data of an insurance company. Actuary (Russian) 5:21--25
\bibitem{Baskakov2010}
Baskakov V, Baskakov I (2010) On ratemaking and other tasks in non-life insurance. Actuary (Russian) 4:37--41
\bibitem{Baskakov1996}
Baskakov V (1996) On an analog of empirical distribution for multivariate censored data. J Math
Sci 81(4):2779--2785
\bibitem{Benjamin1987}
Benjamin B (1977)  General Insurance, Heinemann, London
\bibitem{Dempster1977}
Dempster A, Laird N, Rubin D (1977) Maximum likelihood estimation from incomplete data.  J R
Stat Soc, Ser B 39:1--38
\bibitem{Klein2003}
Klein JP, Moeschberger ML (2003) Survival analysis: techniques for censored and truncated data.
Springer, New York, p 536
\bibitem{Schmidt2008}
Schmidt KD, Zocher M (2008) The Bornhuetter-Ferguson principle. Variance Advancing the Science of Risk 2(1):85--110




\end{thebibliography}
\end{document}